\documentclass[10pt,a4paper]{article}
\usepackage[utf8]{inputenc}
\usepackage[english]{babel}
\usepackage{amsmath}
\usepackage{amsfonts}
\usepackage{amssymb}
\usepackage{graphicx}
\usepackage{lmodern}
\usepackage{url}
\usepackage{authblk}

\usepackage[left=1.25cm,right=1.25cm,top=2cm,bottom=2cm]{geometry}
\begin{document}
	
	\author[1,2]{Francisco Gonzalez Montoya}
	
	\author[3]{Christof Jung}
	
	\author[3]{Thomas H Seligman}
	
	\affil[1]{ School of Chemistry, University of Leeds, Leeds, LS2 9JT, United Kingdom.}
	\affil[2]{Centro Internacional de Ciencias AC - UNAM, Avenida Universidad 1001, UAEM, 62210 Cuernavaca, Morelos, M\'exico.}
	\affil[3]{Instituto de Ciencias F\'isicas, Universidad Nacional Aut\'onoma de M\'exico,
		Av. Universidad s/n, 62251 Cuernavaca, M\'exico}
	
	\title{A dynamical Interpretation of Sequential Decay in Reactive Scattering}
	
	\maketitle 
	
	\begin{abstract}
		The topic of this article is a dynamical explanation of the sequential
		decay in rearrangement scattering. The essential observation is the
		behaviour of trajectories close to the basin boundary of the breakup
		channel. As a most simplistic example of demonstration, we use a version
		of the perturbed three particle Calogero--Moser system in a 1-dimensional
		position space.
	\end{abstract}

	\section{ Introduction}
	
	In reactive scattering, the following possibility has been well known for a long time. 
	A projectile collides with a compound target resulting in some fragment
	leaving the interaction region immediately and a remaining compound fragment
	which decays after some time, which can be large compared to the first
	stage of the reaction. In the end, we have three or even more fragments
	in the final asymptotic state of the scattering event. Such multistep processes
	are known under the name sequential decay. In the present article, we relate these
	events to some newer developments in dynamical system theory in general and
	chaotic scattering in particular. More specifically,
	the purpose is to relate the phenomenon of sequential
	decay to the properties of the basin boundaries of the various arrangement
	channels of the system. This will lead in the end to a dynamical explanation. 
	
	The motivation of this study was in part the importance of sequential decay in 
	nuclear physics, which has its origin in the decay chains of 
	radioactive elements. In the analysis of nuclear reaction data and molecular 
	breakup sequential decay is a frequent phenomenon. 
	Goldansky's 1960 paper on double proton decay \cite{goldansky} is an instructive example 
	of such work, while \cite{nitzan} would be an early reference for sequential
	decay in molecular physics. Note that the sequential decay in the nuclear three
	body problem is not very relevant due to the absence of resonances in these systems.
	Yet recent experiments with Boromean halo states, which we discuss below, appear with
	a strong di-neutron signature. 
	
	It turns out that in many nuclear reactions on light nuclei a
	clear separation of direct reactions and compound reactions 
	can be made and in these cases, the decay of the compound nucleus
	usually can be described as sequential. In heavier nuclei precompound 
	reactions can perturb the picture, but it still largely
	holds. In nuclear fission, the case of spallation obviously is an
	example where decay is not sequential, and in high energy reactions, 
	we may see explosions of the nucleus. Nevertheless, sequential decays 
	will occur frequently even where they are not dominant. 
	The corresponding literature is vast with
	a recent surge of interest in Hoyle states and related alpha
	particle models for alpha like nuclei
	\cite{hafstand,wheeler,medez,qing,freer,ishikawa,dell,rawlinson,cardella}.
	Recent experiments go beyond this in the sense that
	Borromean halo states open a new field where sequential
	decay products can be relevant \cite{marques,bagchi}
	
	The sequential breakup of molecules has recently attracted much attention as
	the excitation with femtosecond laser pulses has become feasible. The dissociation
	of CO$_2$ may be the nearest we have to our toy model, though there is still 
	a significant difference in the fact that the model has its three particles in 
	one dimension rather than in three \cite{maul,wu1}.

	Now the purpose of the present article is to give a dynamical explanation of
	sequential decay with the help of an appropriate model system in the similar 
	spirit that other reaction dynamics in chemistry have been explained with the 
	help of classical models, see for example \cite{GW19,GW20}.
	To keep all explanations on a level as simple as possible we try to construct the
	most minimalistic model system which shows sequential decay. As has become clear 
	from the initial paragraph of this introduction we need at least 3 particles to
	have the possibility of a final arrangement channel with 3 fragments. To keep the
	configuration space as simple as possible, we study a 3 particle model in a 
	1-dimensional position space, i.e. we look at collinear scattering. After a 
	separation of the centre of mass motion, the total system is reduced to a 2 degrees of freedom system, where all important structures in the configuration space and also 
	most explanations in the phase space can be illustrated by 2-dimensional plots.  
	
	For the 2 particle interaction potentials between the various pairs of particles 
	we take inverse squares of hyperbolic functions of position coordinates. These
	potential functions have the great advantage that most quantities in the
	asymptotic dynamics can be expressed in closed form by elementary functions.
	This holds in particular for the transformation to action and angle variables
	for the internal dynamics of 2 particle fragments. The model is constructed
	and explained in all the details in section 2. 
	
	It will turn out that sequential decay occurs for a small layer of initial
	conditions belonging to the basin of the 3 fragment break up channel and lying close to
	the basin boundary of the break up channel. Accordingly in section 3
	we will concentrate on the basin boundaries and the trajectories running close
	to these basin boundaries. Frequently we apply procedures 
	which have been developed before in a previous investigation of the break 
	up channel, see \cite{zj}. Section 4 gives conclusions and some final remarks.
	
	\section{Perturbed Calogero-Moser system}
	
	\subsection{The perturbed three particle Calogero-Moser system with hyperbolic potentials}
	
	As an example of demonstration, we use the version of the
	perturbed Calogero-Moser system with hyperbolic 2-particle
	potentials. See references \cite{cal1982, js} for more details 
	about the properties of the model and its phase space analysis.
	Let us assume 3 particles called $A, B,$ and $C$ moving along a
	1-dimensional position space, i.e. we study collinear scattering.
	The masses of the particles are equal to 1 and the position coordinates
	of the particles are $q_A$, $q_B$ and $q_C$ respectively.
	The conjugate momenta are $p_A$, $p_B$ and $p_C$ respectively.
	We assume that the particle pair $A$ and $B$, and also the particle
	pair $B$ and $C$ have an attractive 2-particle interaction, whereas
	the particle pair $A$ and $C$ has a repulsive 2-particle interaction.
	
	For the total Hamiltonian we take the model system
	
	\begin{equation}
		H = \frac{1}{2}(p_A^2+p_B^2+p_C^2) - \frac{D_A}{\cosh^2{(q_B-q_C)}} + 
		\frac{D_B} {\sinh^2{(q_C-q_A)}} - \frac{D_C}{\cosh^2{(q_A-q_B)}}.
	\end{equation}
	
	We choose $D_B$ as the unit of energy and thereby $D_B$ takes the numerical
	value 1. Next, we assume that the particles $A$ and $C$ are equal and therefore
	$D_A$ and $D_C$ have the same numerical value $D$.
	That is, we will study the case $D_B = 1 \ne D_A = D_C = D$ in the following,
	where only positive values of $D$ make sense.
	
	The figures 1, 2, 3 show examples of the scattering process where the positions of
	the three particles ( horizontal axis ) are plotted as a function of time ( vertical
	axis ). The particles $A$, $B$, and $C$ are represented by the blue, red, and green curve
	respectively. In Fig. 1 the three particles enter as free particles and in the
	long run also leave as free particles.
	In Fig. 2 the free particle $A$ hits a bound state of the
	particles $B$ and $C$ and a reaction occurs where in the end the free particle $C$
	moves away from a bound state of the particles $A$ and $B$. In Fig. 3  the free
	particle $C$ hits a bound state of the particles $A$ and $B$. This time the final
	state of the reaction are three free particles.
	\begin{figure}[h!]
		\begin{center}
			\includegraphics[scale=0.5]{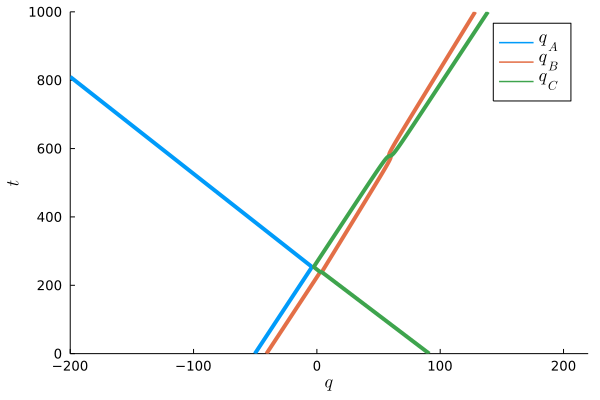}
			\caption{Position of the three particles vs time. The system starts with three 
				free particles in the incoming asymptotic region, then the particles interact close 
				to $q=0$ and the system finishes again as three free particles.  
				The potential is the perturbed Callogero-Moser potential from Eq. 1 for the nonintegrable case with $D_B=1$,  $ D_A = D_C =$ 0.9. We have the total energy $E$=0.1.  The numerical calculation is done using the Taylor integrator order 25 implemented in \cite{perez} to keep the error in the energy small in all the trajectories, particularly when the particles are close and the standard numerical methods fail. \label{fig:00} }
		\end{center}
	\end{figure}
	\begin{figure}[h!]
		\begin{center}
			\includegraphics[scale=0.5]{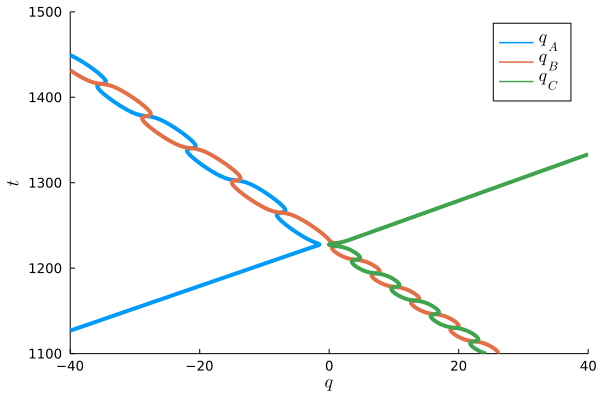}
			\caption{ Position of the three particles vs time. The system starts as the free particle $A$ and
				a bound state of the particles $C$ and $B$ and ends as a free particle $C$ and a bound state of the 
				particles $A$ and $B$. The potential parameters have the same values as in  Fig. \ref{fig:00} . \label{fig:AB}}
		\end{center}
	\end{figure}
	\begin{figure}[h!]
		\begin{center}
			\includegraphics[scale=0.5]{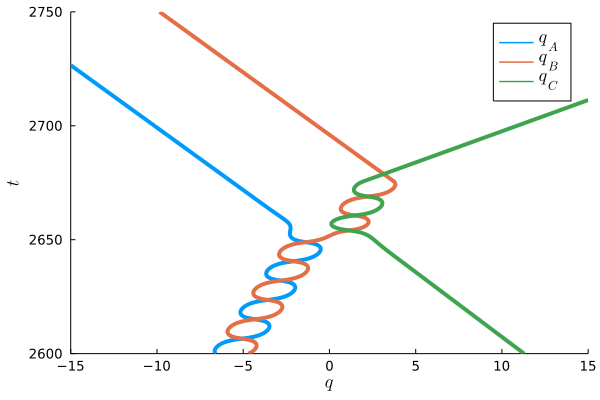}
			\caption{ Position of the three particles vs time. The system starts again as a free particle $C$
				and a bound state of the particles $A$ and $B$. However this time it ends as three free particles.
				The parameter values of the potential are the same as in Fig. \ref{fig:00}. \label{fig:A0} }
		\end{center}
	\end{figure}

	\newpage

	In order to get rid of the centre of mass motion we transform to the 
	centre of mass coordinate $S$ and the relative coordinates $R$ and $r$
	defined as
	
	\begin{eqnarray}
		S &=& (q_A + q_B + q_C) / 3,  \nonumber \\
		R &=& \sqrt{1/6} (q_A+q_C - 2 q_B), \nonumber  \\
		r &=& \sqrt{1/2} (q_C - q_A)   \nonumber \\
	\end{eqnarray}
	
	and the corresponding conjugate momenta
	
	\begin{eqnarray}
		p_S &=& p_A + p_B + p_C ,   \nonumber \\
		p_R &=& \sqrt{1/6} ( p_A + p_C - 2 p_B ) ,\nonumber   \\
		p_r &=& \sqrt{1/2} (p_C - p_A) . \nonumber \\
	\end{eqnarray}
	
	From now on we do no longer care about the centre of mass degree of freedom (dof)
	and set $S$ and $P_S$ permanently to the value 0. In the new coordinates the 
	Hamiltonian for the two relative degrees of freedom has the form
	
	\begin{equation}
		H = \frac{p_R^2}{2} + \frac{p_r^2}{2} + V(R,r),
	\end{equation}
	
	where the potential energy $V(R,r)$ is defined as
	
	\begin{equation}
		V(R,r) = - \frac{D}{\cosh^2{(\sqrt{3/2}R+ \sqrt{1/2}r)}} 
		+ \frac{1}{\sinh^2{(\sqrt{2} r)}} - \frac{D}{\cosh^2{( \sqrt{3/2}R - \sqrt{1/2} r)}}. \nonumber
	\end{equation}
	
	Up to an unusual front factor, the new coordinates are the Jacobi coordinates
	as seen from the particle $B$. The front factors have been included in order
	to transform the reduced masses to the value 1. Remember that the transformation
	$p \to \sqrt{m} p$, $ q \to q / \sqrt{m}$ is a canonical transformation.
	
\begin{figure}[h!]
		\begin{center}
			\includegraphics[scale=0.5]{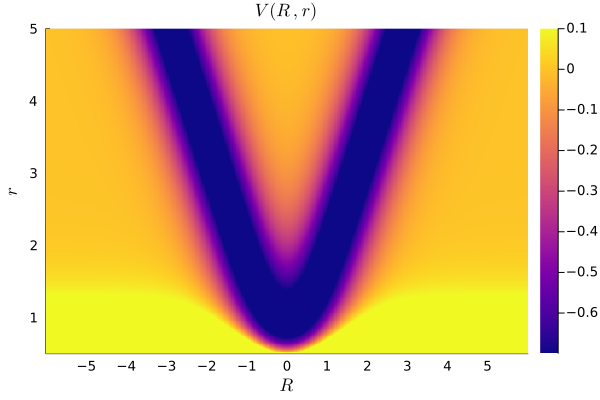}		
			\caption{Potential energy $V(R,r)$ in colour scale as a function of the relative 
				coordinates $R$ and $r$. The values of the parameters are $D_B=1$,  $ D_A = D_C = 0.9$.
				The potential energy surface has a valley plotted in dark colour. 
				In the asymptotic region, the branch of the valley 
				on the right corresponds to the scattering channel $C$ and the other branch 
				on the left to the channel $A$. The potential energy goes to infinity on the line $r$=0 and 
				is almost flat outside the valley and for $r$ large. 
				\label{fig:potential_energy}}
		\end{center}
	\end{figure}

	A plot of the total potential is given in Fig. \ref{fig:potential_energy}. Between the 
	particles $A$ and $C$ we have a repulsion which becomes infinite for $q_A$ approaching $q_C$.
	Therefore we can restrict all further considerations to the case that particle
	$C$ is to the right of particle $A$, i.e. we can restrict the investigation to
	the half plane $r > 0$ of the configuration space of the relative coordinates.
	The total potential shows a curved valley whose bottom line converges to the 
	straight line $L_C$ given by $r = \sqrt{3}R$ for $R \to + \infty$ and to the 
	straight line $L_A$ given by $r = - \sqrt{3}R$ for $R \to - \infty$.

	Far away from these two straight lines and also far away from the $r$ axis
	the potential goes to zero exponentially fast. In Fig. 4 we can appreciate
	well how the potential converges to straight potential channels for large
	values of the coordinate $r$. In the central
	region where all three potential contributions are important, we see how the
	potential valley makes a curve in this interaction region.
	
	
	The simplest and shortest periodic orbits always play an important role in the
	dynamics. In our case, it is clear by symmetry arguments that for a negative
	value of the total energy there exists a simple periodic orbit $\gamma$
	oscillating along the $r$ axis in the potential valley.
	
	The bottom line of the potential valley has a 1:1 projection on the $R$ axis and later we call $v$ the
	coordinate along the potential valley where between $v$ and $R$ a 1:1 relation
	holds. That is, $v$ is the reaction
	coordinate along the valley. We call $u$ the coordinate perpendicular to $v$ and
	pointing in the positive $r$ direction.
	
	The part of the valley for $R$ large and positive will be called the potential
	channel $C$. And the part of the valley for $R$ large and negative will be
	called the potential channel $A$.
	
	Fig. 5 gives the 3 trajectories shown before in Figs. 1, 2, 3 again but
	this time plotted as curves in the space of the relative coordinates. The blue
	curve is the trajectory from Fig. 1, the green curve is the trajectory from Fig. 2
	and the orange curve is the trajectory from Fig. 3.
	
	
	\begin{figure}[h!]
		\begin{center}
			\includegraphics[scale=0.5]{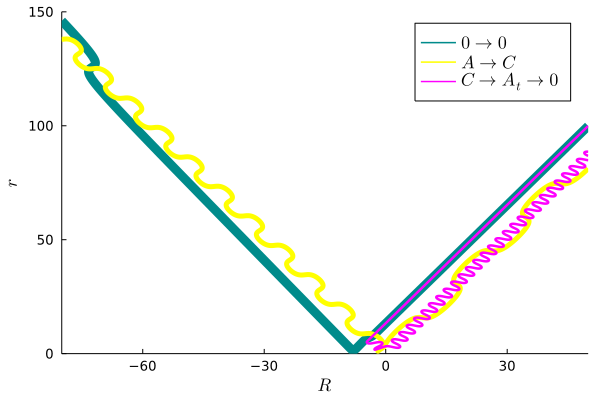}
			\caption{ The trajectories plotted in relative coordinates $R$ and $r$ which have already 
				been shown before in 3 particle coordinates in Figs. \ref{fig:00},\ref{fig:AB},\ref{fig:A0}. We can 
				appreciate how the potential energy in Fig. \ref{fig:potential_energy} defines the scattering channels.
				The trajectory that goes from channel $C$ to channel $0$, corresponding to Fig. \ref{fig:A0}, 
				has a transient behaviour close to the origin. This transient motion is represented by the segment of the 
				trajectory with $R<0$.\label{fig:trajectories_chanels}}
		\end{center}
	\end{figure}
	
	\subsection{ The asymptotic arrangement channels and their asymptotic labels}
	
	In a system consisting of 3 particles we have various possibilities to group 
	these particles into fragments and form bound states. First, for $D>1$ and for a total 
	energy $E$ smaller than the asymptotic depth $-D$ of the potential valley, 
	i.e. for $E<-D<-1$, only 3 particle bound states exit. 
	They are not relevant for the topic of the present article. Second, there
	is the possibility of a bound state of the particles $B$ and $C$ bound together
	and the particle $A$ moving freely and far away from $B$ and $C$. Such states will
	be called states in the arrangement channel $A$ and their trajectories move in
	the channel $A$ of the potential. Next, there are states where the particles
	$A$ and $B$ are bound together and the particle $C$ moves freely and far away from the
	particles $A$ and $B$. Such states are called arrangement $C$ and their corresponding
	trajectories move in the channel $C$ of the potential. In our particular model,
	the particles $A$ and $C$ have a repulsive interaction. Therefore an arrangement $B$
	does not exist, where particle $B$ would move freely relative to a bound state
	of the particles $A$ and $C$. Finally, there exists the breakup channel, also
	called arrangement channel 0, where all 3 particles move freely and asymptotically any particle
	moves far away from the other particles. They correspond to trajectories in the
	position space of relative coordinates far away from the potential valley and
	also far away from the $R$ axis, i.e. in the region where the tails of all
	potentials are exponentially small.
	
	First, let us look at the case $D>1$ where there exists on the $r$ axis a local 
	minimum of the total potential with energy $E_m<-D$ and we have only 3 particle 
	bound states in the energy interval $(E_m, -D)$. For total energies $E$ in the 
	interval $(-D,0)$ the trajectories belonging to the chaotic invariant set 
	form 3 particle bound states. Trajectories running along the stable manifolds 
	of localized subsets can start in the initial arrangement $A$ or $C$ and they 
	end as 3 particle bound states. However, such trajectories only occupy a subset
	of measure 0 in the energy shell. Trajectories running on unstable manifolds
	of localized subsets show the corresponding time reversed behaviour. Stable
	3 particle bound states can exist in KAM islands around stable periodic orbits.
	The rest of the phase space is occupied by trajectories leaving along the 
	potential valley in the past and also in the future. 
	
	For $D<1$ there is a saddle point of the potential along the $r$ axis with 
	saddle energy $E_s>-D$ and the periodic orbit $\gamma$ is the only 3 particle 
	bound state. For an energy in the interval $(-D, E_s)$ trajectories starting
	in channel $A$ return to channel $A$ and trajectories starting in channel $C$ return
	to channel $C$. For energies in the interval $(E_s, 0)$ transitions from channel
	$A$ to channel $C$ and vice versa become possible. Here trajectories running on
	the unstable or the stable manifolds of the periodic orbit $\gamma$ are 
	exceptions of measure 0 which start or end as 3 particle bound states respectively.
	
	For energies $E>0$ also the channel 0 is energetically accessible. That is, for a positive
	total energy trajectories starting in the channels $A$ or $C$ can end in channel 0
	by breakup processes. For very well selected initial conditions also the reverse
	can happen, i.e. trajectories starting in the channel 0 can end in one of the
	2-fragment channels. However, such processes need a very delicate preparation
	of the initial conditions and are of little practical importance. Some of these
	possibilities have already been illustrated by the Figs. 1, 2, 3, 5.
	
	For our particular case of study there are no localized orbits
	for positive energies and therefore also no stable or unstable manifolds of
	localized subsets. As a consequence for $E>0$ we also do not have trajectories
	which start or end as 3-particle bound states. 
	
	To describe the motion in the asymptotic channels $A$ and $C$ we use the coordinates
	$u$ and $v$ already mentioned before. They are defined in terms of the coordinates
	$R$ and $r$ in the channel $A$ as
	
	\begin{eqnarray}
		u &=& \sqrt{3} R/2 + r/2 , \nonumber \\
		v &=&  R/2 - \sqrt{3} r/2 \nonumber \\
	\end{eqnarray}
	
	and in channel $C$ as
	
	\begin{eqnarray}
		u &=& - \sqrt{3} R/2 + r/2,  \nonumber \\
		v &=& R/2 + \sqrt{3} r/2.  \nonumber \\
	\end{eqnarray}
	
	Asymptotically in the channels $A$ and $C$ the motions in longitudinal 
	and in transversal direction separate.
	The asymptotic Hamiltonian has in both cases the functional form
	
	\begin{equation}
		H_{as} = \frac{p_v^2}{2} + H_{\perp}(u,p_u),
	\end{equation}
	
	where
	
	\begin{equation}
		H_{\perp}(u,p_u) = \frac{p_u^2}{2} - \frac{D}{\cosh^2{(\sqrt{2} u)}}.
	\end{equation}
	
	For negative values of the transverse energy, i.e. negative values of the Hamiltonian from
	Eq.8, it is convenient to transform the transverse motion ( the $u$ degree of freedom ) to
	action and angle variables $I, \phi$. For the $cosh^{-2}$ potential this transformation is
	well known in terms of elementary functions and here we only give the result
	
	\begin{equation}
		I(u,p_u) = \sqrt{D} - \sqrt{-E_{\perp}},                 
	\end{equation}
	
	\begin{equation}
		\phi(u,p_u) = \arcsin [ \sqrt{-E_{\perp}/(E_{\perp}+D)} \sinh(\sqrt{2}u)],
	\end{equation}
	
	where in Eqs. 9 and 10 for $E_{\perp}$ the Hamiltonian from Eq.8 has to be inserted. 
	Inversion of Eq.9 gives
	
	\begin{equation}
		H_{\perp}(I, \phi) = - ( I - \sqrt{D} )^2.
	\end{equation}
	
	Note that the transformation to action and angle variables only makes sense when $E_{\perp}$,
	the energy in the transverse $u$ degree of freedom is smaller than 0.
	As it must be for a 1-dof system, $H_{\perp}$ is independent of the angle and accordingly
	$I$ is an asymptotically conserved quantity. The velocity $\omega$ of the angle $\phi$ is given as
	
	\begin{equation}
		\omega = \frac{\partial H_{\perp}(I, \phi)} { \partial I}  = -2 ( I - \sqrt{D} ).
	\end{equation}
	
	Next, we have to decide how we label asymptotes in the channels $A$ and $C$. As in any Hamiltonian
	system we have as a first label the value of the total energy $E$. Second, we have the
	distribution of the energy between the longitudinal degree of freedom $v$ and the 
	transversal degree of freedom $u$. This can be done by giving either the value of the
	longitudinal momentum $p_v$ ( equivalent to giving the longitudinal kinetic energy ) 
	or equivalently by giving the transversal action $I$ ( equivalent to the transversal energy ).
	Finally, we need a quantity which gives the relative phase shift between the longitudinal and
	the transversal motion. This is a reduced angle variable which is constant in the
	asymptotic region. We can use the reduced angle variable  $\phi_{red}$ defined as
	
	\begin{equation}
		\phi_{red} = \phi - \omega \frac{v}{p_v}.
	\end{equation}
	
	Note that the 3 asymptotic labels $E$, $I$ and $\phi_{red}$ are constant along the
	asymptotic motion in the 2-fragment channels. Therefore they are appropriate to
	label initial or final asymptotes. 
	
	Of course, these label are not appropriate for
	asymptotes in channel 0. In the breakup channel asymptotic trajectories are straight
	lines whereas the first two labels we can use either the total energy and the direction
	of the momentum $\theta$ or equivalently the vector momentum. As a third label we use either
	the impact parameter relative to the origin or equivalently the angular momentum of
	the trajectory. Note that also these 3 asymptotic labels are constant along asymptotic
	trajectories of the channel 0. Thereby the asymptotic motion is superintegrable in
	any case and in any asymptotic channel.
	
	\subsection{Integrability properties}
	
	The 2 particle potentials all depend on differences between the particle
	coordinates only. Therefore the total momentum of the system $p_S$ is a conserved
	quantity and we have already separated away the centre of mass motion by a transition
	to relative coordinates. As in all Hamiltonian systems, the Hamiltonian
	function is a second conserved quantity. And if all 3 potential strengths 
	$D_A$, $D_B$ and $D_C$ have the same numerical value ( which we can set to 
	the value 1 in this case ) then we even have a third conserved quantity $K$
	in closed form in terms of elementary functions.
	
	\begin{equation}
		K = p_A p_B p_C + \frac{p_A}{ \cosh^2{(q_B-q_C)}} - \frac{p_B}{\sinh^2{(q_A-q_C)}} +
		\frac{p_C}{\cosh^2{(q_A-q_B)}}.
	\end{equation}
	
	For a detailed discussion of this very particular case see \cite{cal1982}.
	
	If the 3 coupling parameters $D_A$, $D_B$, and $D_C$ are not equal then the
	conserved quantity $K$ does not exist any longer. In the case studied in this
	article, namely $D_B=1$, $D_A=D_C=D$ the properties of chaos are as follows.
	For $D>1$ and $-D<E<0$ we find a chaotic invariant set in the phase space, a chaotic saddle.
	In a Poincar\'e map with the intersection condition $u=0$ with a specific intersection
	orientation it is represented by a ternary symmetric horseshoe where
	the outer fixed points sit at $v = \pm \infty, p_v =0 $ and the central fixed
	point sits at $v=0, p_v=0$. In this chaotic case, general scattering trajectories 
	show transient chaos and the scattering functions show a fractal set of 
	singularities as it is usual for chaotic scattering. For general information
	on transient chaos and its relationship to chaotic scattering see \cite{tel2} 
	and chapter 6 of \cite{tel}. For publications on chaotic scattering with an
	explanation of many concepts which we will mention later briefly without any
	detailed explanation see also \cite{san,ffc} and chapter 4 in \cite{reichl}.
	For the intriguing relation between the integrability of the Hamiltonian and 
	the integrability of the $S$-matrix see \cite{js}.
	
	For $D<1$ the only localized trajectory for $E<0$ is the periodic orbit $\gamma$ 
	oscillating along the $r$ axis. Accordingly, there is no chaotic saddle and no 
	scattering chaos. In this case, a third conserved quantity could be constructed 
	by the transport of any asymptotic initial condition over the subset $S$ of the phase space 
	consisting of the complete phase space without points lying on $\gamma$ or its 
	stable and unstable manifolds. For example, to any point $x \in S$ we construct the
	trajectory through $x$ and assign to $x$ the value of $I$ on the initial asymptote
	of this trajectory. This quantity is constant along any trajectory by construction
	and it defines a smooth function on the subset $S$. Note that in the presence 
	of a chaotic saddle with its fractal geometry this procedure would not result in
	a piecewise smooth function on the phase space.
	
	For $E>0$ this construction also works for all points belonging to trajectories
	with an incoming asymptote in one of the 2 fragment channels. For points lying on
	trajectories with an initial asymptote in channel 0 we can use instead the 
	initial direction of the momentum $\theta$ and assign it to all points along the trajectory.
	This procedure constructs a piecewise smooth function on the phase space. Note,
	that for $E>0$ no localized trajectories exist. Therefore there is no need to
	take away periodic orbits and their stable and unstable manifolds from the domain of this
	construction. The only exceptional subset is the boundary of the basin of the
	channel 0 which will be studied in the next section.

	\section{ Basin boundary of the breakup channel}
	
	We can divide all points $x$ of the phase space into various subsets according
	to the channel in which the trajectory through this point $x$ ends in the far future.
	The basins of the channels $A$, $C$, and 0 are all such initial points $x$ whose trajectories
	end up in channel $A$, $C$ or 0 respectively. These basins will be called $B_A$, $B_C$
	and $B_0$ respectively. Initial points lying on stable manifolds of localized subsets
	or on the localized subsets themselves do not end in any one of the asymptotic 
	channels $A$, $C$ or 0. Therefore they do not belong to any one of the basins $B_A$, 
	$B_C$ or $B_0$ respectively. We call this subset $B_L$.
	
	Usually, the boundaries between the various basins are either stable manifolds of
	localized subsets or they are stable manifolds of particular subsets of final
	asymptotes. For a negative total energy $E$ and $D>1$ the phase space contains a chaotic
	saddle and its stable manifolds are the boundary between the channels $A$ and $C$.
	For a chaotic saddle, its stable manifolds form a fractal and therefore also the basins
	and their boundaries are fractal. 
	
	For $D<1$ and $E \in (-D, E_s)$ the basins of the channels $A$ and $C$ are well separated
	in the position space and do not have any common boundary, they are disjoint components
	of the energy shell. In this case, the basin boundaries consist of energetic boundaries 
	only. For $D<1$ and $E \in (E_s, 0)$ the energy shell has a single component only
	which is divided into $B_A$, $B_C$ and $B_L$. The basin boundary is formed by the stable
	manifold of the periodic orbit $\gamma$.
	
	For a positive total energy, the case studied in the present work, 
	we do not have any localized trajectories. Therefore,
	there can not exist any stable manifolds of localized subsets of trajectories.
	This implies the absence of any chaotic saddle and of any fractal basin boundaries.
	In addition, there are no direct boundaries between $B_A$ and $B_C$. When we deform
	a trajectory ending in channel $A$ smoothly into a trajectory ending in channel $C$
	then the deformation process must pass through a set of trajectories ending in
	channel 0. Accordingly, the only boundaries which exist are boundaries between 
	the basin $B_A$ and $B_0$ and boundaries between the basins $B_C$ and $B_0$. Because of the
	absence of chaos for positive $E$ these boundaries can not be fractal. Compare the
	analogous arguments for the model system explained in \cite{zj}. Because of the
	discrete symmetry of the system, it is sufficient to study in the rest of this
	section trajectories starting in the channel $A$ and to investigate in detail the boundary 
	between the basins $B_A$ or $B_C$ and $B_0$.
	
	\subsection{ Approach to the basin boundary from the side of the 2-fragment channels}
	
	In the next 2 subsections, we describe the behaviour of typical 
	trajectories when their initial conditions approach the basin boundary. First
	we do it for the case when the initial condition approaches this boundary
	from the side of the 2 fragment channel $A$. If the trajectory ends in a 2 fragment
	channel then the final asymptote moves along the $v$ coordinate with constant speed.
	Simultaneously it oscillates in the $u$ direction with a frequency given by Eq. 12.
	The energy in the $u$ degree of freedom ( $E_\perp$ given by Eq. 11 ) is negative 
	and approaches 0 from below when the initial condition approaches the basin
	boundary from the side of the 2 fragment channel. In the limit of the distance
	to the boundary going to zero the frequency from Eq. 12 also goes to zero, i.e. the
	oscillation period $T=2 \pi / \omega$ goes to infinity. If the trajectory moves 
	in the $v$ direction ( i.e. along the axis of the potential channel from Fig. 1 ) 
	with momentum $p_v$ then the oscillation wavelength of the trajectory along the
	potential channel is $\lambda = p_v T $ and also goes to infinity when the
	initial condition ( and therefore also the whole trajectory ) approaches the
	basin boundary. In the limit of running exactly on the basin boundary, 
	graphically at very large longitudinal distance $v$ the trajectory seems to 
	run parallel to the potential valley at a finite transverse distance.
	
	It is common in scattering experiments that we prepare the total energy and also
	prepare the oscillation energy of the two particle fragments, in our case given
	by the oscillation action from Eq. 9. In contrast usually we do not have control
	over the reduced phase of the oscillation of the two particle fragment, in our
	case given by Eq. 13. We assume now that this reduced phase has a random distribution
	over the interval $[0, 2 \pi ]$. To study the trajectories close to the basin
	boundary we do the following. We fix in the incoming channel $A$ values of $E=0.01$ 
	and $I =0.91 $, scan $\phi_{red}$ in the interval $[0, 2 \pi]$ and let the trajectories
	run until they reach the outgoing asymptotic region. 
	
	For the very small energy $E=0.01$ used from now on, we have in the whole domain 
	$\phi_{red} \in [0, 2 \pi]$ one interval where the trajectories end in channel $A$, one
	interval where the trajectories end in channel $C$, and in between two very short
	intervals where the trajectories end in channel 0.
	
	Interesting for us is now the graphical presentation of the result in a region
	of $\phi_{red}$ which contains an interval ending in channel 0 and some small
	parts of the adjacent intervals ending in the 2 fragment channels. It is Fig. 6.
	
	
	\begin{figure}
		\begin{center}
			
			(a) \includegraphics[scale=0.6]{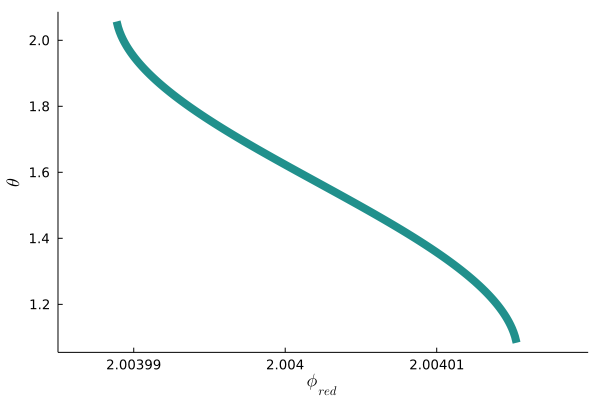}\\
			(b) \includegraphics[scale=0.6]{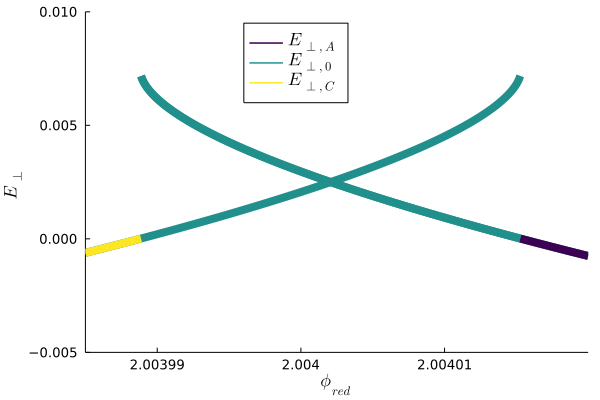}
			
			\caption{ The final angle $\theta$ in the upper panel and the final 
				transverse energies $E_{\perp,A}$ as the blue curve and $E_{\perp,C}$ as the yellow curve  in the lower panel, all as a function of the initial reduced phase $\phi_{red}$. 
				For additional explanations see the main text. }
			
		\end{center}
	\end{figure}

	
	In part (a) of Fig. 6, the outgoing direction $\theta$ of the momentum is shown
	in the $\phi_{red}$ interval which leads to the final channel 0. Remember that this
	quantity $\theta$ only makes sense for the asymptotic channel 0. This angle has
	values between $\pi/3$ and $2 \pi/3$. These limiting values are exactly the directions
	of the straight lines $L_C$ and $L_A$ respectively, i.e. the asymptotic directions
	of the potential valleys for the asymptotic channels $C$ and $A$. 
	This demonstrates graphically how the outgoing straight trajectory of the channel 0 
	becomes parallel to the asymptotic channel axis of the potential
	when the trajectory approaches the boundary of channel $0$. 
	
	Part (b) of Fig. 6 shows the transverse energy of the outgoing asymptotes as
	function of the initial reduced phase $\phi_{red}$. 
	The blue curve gives the final asymptotic energy $E_{\perp,A}$ corresponding to the channel $A$ 
	and the yellow curve gives $E_{\perp,C}$ corresponding to the channel $C$ . 
	For the trajectories that ends in channel $0$, we can use both definition 
	of the transverse energy of the outgoing asymptotes to define $E_{\perp,0}$.
	A trajectory ends in channel $A$ exactly when $E_{\perp,A} < 0$ and it ends in channel $C$ exactly when
	$E_{\perp,C}<0$.  The boundary point between basins $B_A$ and $B_0$ is the point where
	$E_{\perp,A} = 0$ and the boundary point between basins $B_C$ and $B_0$ is the point where
	$E_{\perp,C} = 0$.
	
	\newpage
	
	In the next two figures, we show some trajectories in the position space running close
	to the basin boundary of channel 0. The Fig. 7 presents 6 trajectories close to the
	boundary between basins $B_A$ and $B_0$ and Fig. 8 presents 6 trajectories close to the boundary
	between basins $B_C$ and $B_0$. For each one of these figures there are three trajectories 
	from the basin of the two fragment channel ( panels (a), (b) and (c)) and three trajectories 
	from the basin of the channel 0 ( panels (d), (e) and (f)) respectively. 
	
	\begin{figure}
		\begin{center}
			
			(a)\includegraphics[scale=0.4125]{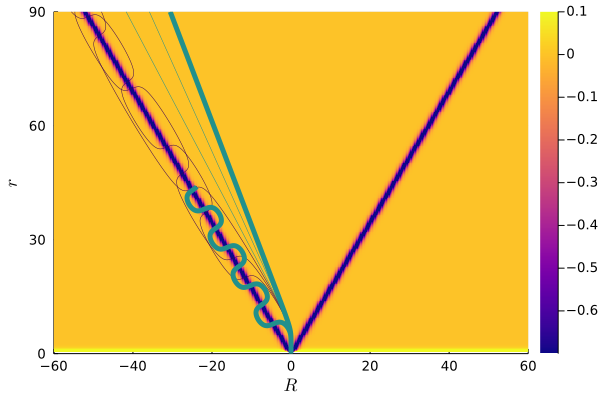}
			(d)\includegraphics[scale=0.4125]{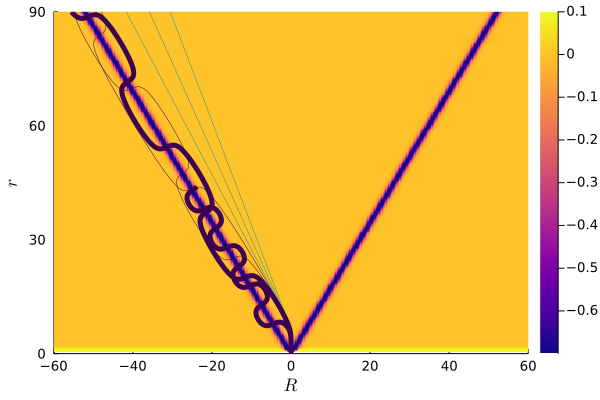}\\
			
			(b)\includegraphics[scale=0.4125]{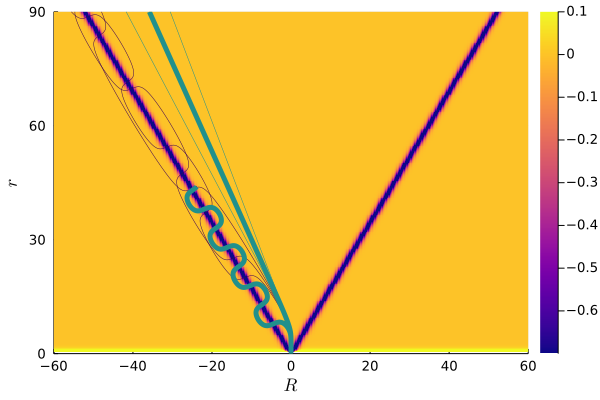}
			(e)\includegraphics[scale=0.4125]{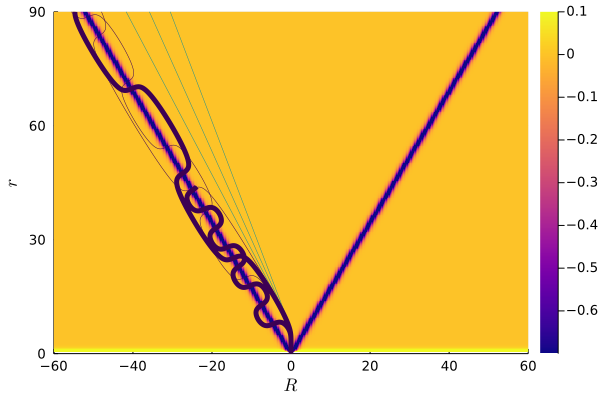}\\
			
			(c)\includegraphics[scale=0.4125]{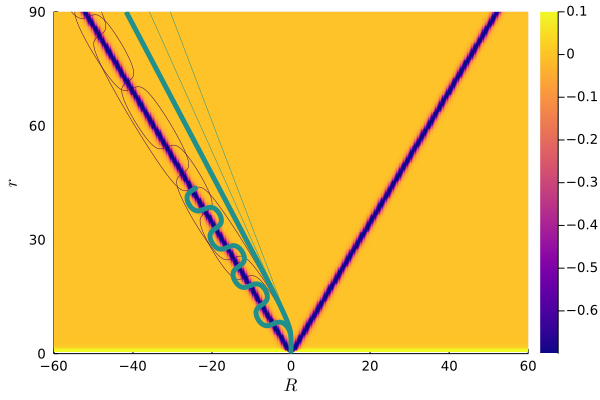}
			(f)\includegraphics[scale=0.4125]{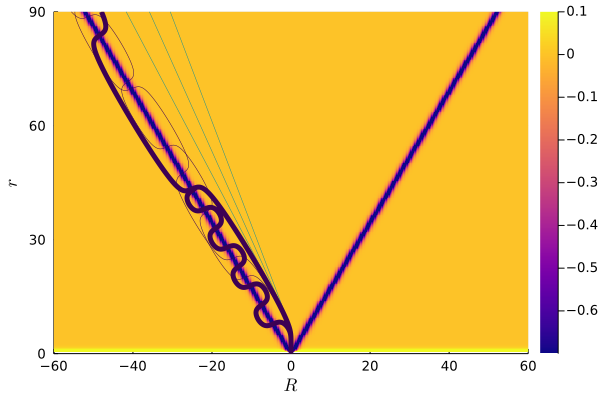}
			
			\caption{ Trajectories close to the boundary between final asymptotic channels $A$ and $0$. $E$ = 0.01, $D$ =0.9.
				The trajectories highlighted on the left panels end in the channel $0$ 
				and the trajectories highlighted on the right panels end in the channel $A$. 
				The trajectories highlighted on lower panels are close to the boundary between channel $A$ and 0. 
				\label{fig:trajectories_b1}}
		\end{center}
	\end{figure}
	
	\begin{figure}
		\begin{center}
			
			(a)\includegraphics[scale=0.4125]{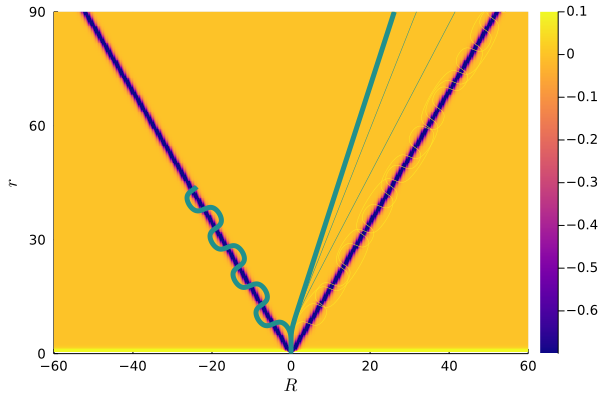}
			(d)\includegraphics[scale=0.4125]{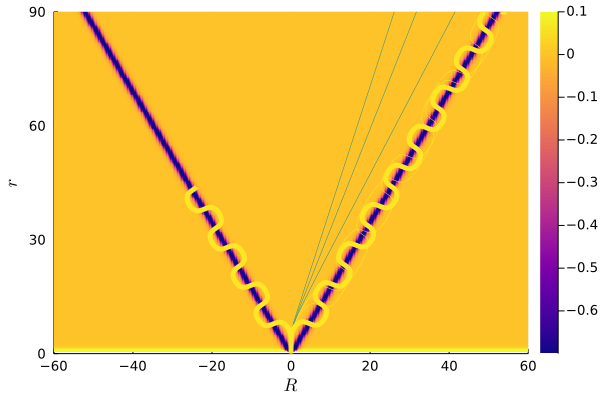}
			
			(b)\includegraphics[scale=0.4125]{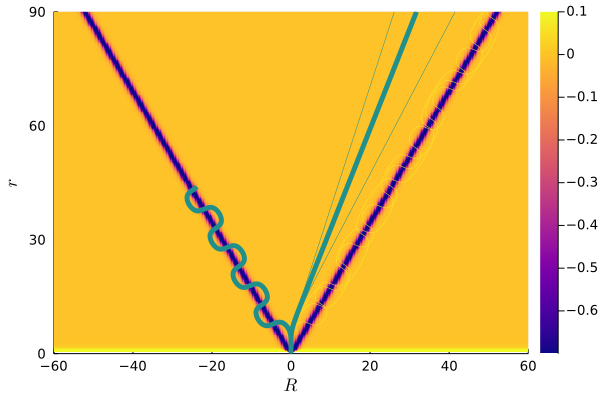}
			(e)\includegraphics[scale=0.4125]{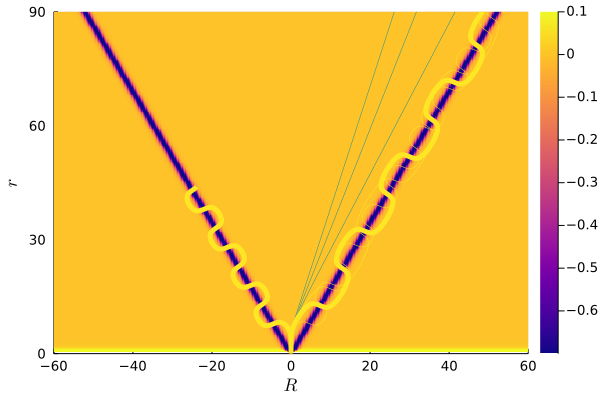}
			
			(c)\includegraphics[scale=0.4125]{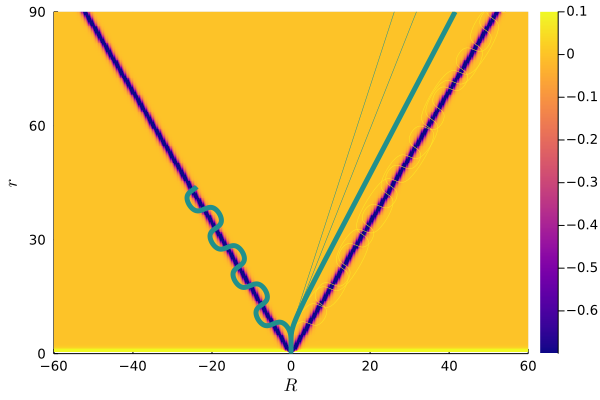}
			(f)\includegraphics[scale=0.4125]{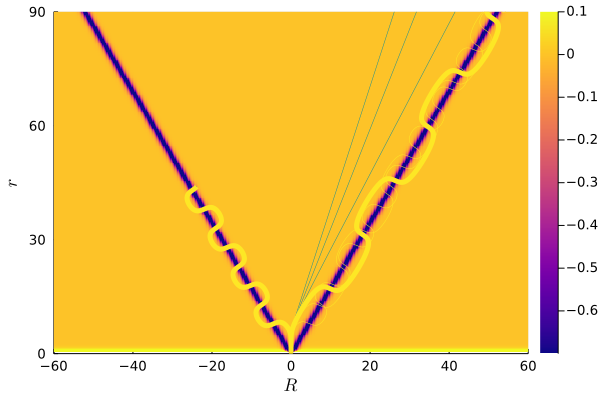}
			
			\caption{ Trajectories close to the boundary between final asymptotic channels $C$ and $0$. 
				$E$ = 0.01, $D$ =0.9.
				The trajectories highlighted on the left panels end in the channel $0$ 
				and the trajectories highlighted on the right panels end in the channel $C$. The trajectories 
				highlighted on lower panels are close to the boundary between channel $C$ and 0.
				\label{fig:trajectories_b2}}
		\end{center}
	\end{figure}
	
	The panels (a), (b), (c) of Fig. 7 show a sequence of 3 trajectories in the position space 
	where the initial condition comes closer to the basin boundary from the side of $B_A$.
	We observe how accordingly the oscillation period $T$ in $u$ direction and the 
	wavelength $\lambda$ along the $v$ direction become larger. Eventually they
	converge to $\infty$ in the limit where the initial condition reaches the basin
	boundary. In the panels (a), (b) and (c) of Fig. 8 we see the corresponding
	behaviour when the initial condition approaches the boundary between $B_C$ and $B_0$
	from the side of $B_C$.
	
	\subsection{ Approach to the basin boundary from the side of the breakup channel}

	Now let us look at trajectories where the initial condition lies close to the basin
	boundary but this time on the side of the 3 fragment channel 0. In this case,
	the energy in the $u$ degree of freedom is small and positive. That is, the
	trajectory leaves the potential valley and in the long run the $u$ coordinate
	becomes large positive or large negative. The relative angle between the asymptotic
	trajectory and the potential valley is slightly different from zero. 
	
	In the sequence of plots of trajectories in the position space in the panels (d), (e) and (f) 
	of Fig. 7 the initial condition comes closer to the basin boundary
	between $B_A$ and $B_0$ from the side of $B_0$.
	We observe how accordingly the asymptotically straight outgoing trajectory
	becomes more and more parallel to the channel axis of channel $A$. 
	At the same time, the last crossing of the axis of the potential channel $A$ is pushed out to
	larger values of the reaction coordinate $v$ when the initial condition
	converges to the boundary point.
	In Fig. 8 the corresponding behaviour is illustrated near the boundary between
	$B_C$ and $B_0$.
	
	Importantly, very close to the boundary point it needs a very long observation time 
	to distinguish such a trajectory ending in the channel 0 with very small positive
	transverse energy from one ending in the channel $A$ with a negative transverse energy 
	( here it is the bound state energy ) very close to 0.
	
	As a result, we have obtained the following dynamical picture of the phenomenon of
	sequential decay. If the initial conditions are close to the boundary of the basin
	$B_0$ with some two fragment channel $X$ ( where in our example of the perturbed
	Calogero-Moser system $X$ can be either $A$ or $C$ ) then in the collision the particle
	$X$ separates rather quickly from the remaining set of particles with an energy
	clearly larger than 0. However, the remaining group of particles has a total
	energy a little larger than 0. Therefore this remaining group is close to be
	bound but not really bound and this remaining group dissolves slowly with very
	small kinetic energy between its fragments. Accordingly, it takes some time to observe clearly and without
	doubt the decay of this remaining group. The observer notices one particle
	( or fragment ) leaving instantaneously and other particles ( or fragments ) 
	separating later with very small relative kinetic energy, a typical sequential decay.
	
	It depends on the observational resolution, where we make the distinction between
	sequential decay and direct decay into three or more fragments. On this
	somewhat arbitrary choice depends the width of the layer in basin $B_0$ along its
	boundary which leads to the observation of sequential decay. As a consequence
	also the relative fraction of scattering events which are classified as
	sequential decay depends on this choice when we shoot in a beam of fragments without any control over the reduced oscillation phase.
	
	
	To give a pictorial impression of the basins for
	the various channels in the set of the initial conditions
	and also a partial picture of the basins in the phase
	space, one can do the following. We select an appropriate
	2-dimensional plane in the set of all initial conditions,
	place a grid of many pixels on this plane and colour
	each one of this pixels according to which basin it
	belongs, i.e. according to the channel in which the
	corresponding trajectory ends in the far future.
	
	In Fig. 9 we choose the $\phi_{red}$--$I$ plane
	for the fixed value of the total energy $E = 0.01$ and
	initial conditions in the channel $A$. Pixels whose
	corresponding trajectories end in channel $C$ and coloured yellow.
	Pixels whose trajectories end in channel $A$ are coloured
	cyan. And pixels whose trajectories end in channel 0
	are coloured dark blue.
	In contrast in Fig. 10, we fix the transverse action
	$I$ at the value 0.91 and choose the $\phi_{red}$--$E$ 
	plane in the initial channel $A$ and apply the same
	procedure as in the previous figure to colour the pixels.
	We observe that for these initial
	conditions most of the trajectories end in channel $C$, i.e.
	they just advance along the potential channel with a
	monotonically increasing reaction coordinate. Only for
	a small fraction of the initial condition the trajectory
	does something more complicated in the interaction region,
	interchanges strongly transversal and longitudinal energy
	and thereby manages to end in channel $A$ or in channel 0.
	
	Basically the domains of the plots are divided into
	layers belonging to the various final channels.
	Partially these layers are extremely thin, thinner
	than the pixel resolution. Therefore partially the stripes belonging
	to the various channels look broken into disjoint pieces,
	whereas in reality they should be continuous stripes going
	around in $\phi_{red}$ direction. This appearance of extremely
	thin layers seems to be a general phenomenon. It has
	also been observed in the Refs. \cite{GAC,KK} in model calculations for
	molecular reactions. The corresponding layers in the
	complete phase space are obtained by the transport of these
	layers in the set of the asymptotic initial conditions
	by the flow.


	\begin{figure}
		\centering
		\includegraphics[scale=0.6]{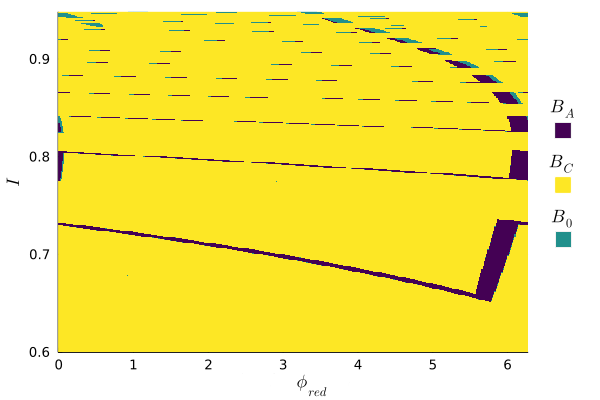}
		\caption{The $\phi_{red}$--$I$ plane for fixed
			$E=0.01$ in the set of initial conditions in channel $A$.
			Each pixel in this plane is coloured according to the
			channel in which the corresponding scattering trajectory
			ends in the far future. Yellow means basin $B_C$,
			cyan means basin $B_A$ and dark blue means basin $B_0$.}
	\end{figure}

	\begin{figure}
		\centering
		\includegraphics[scale=0.6]{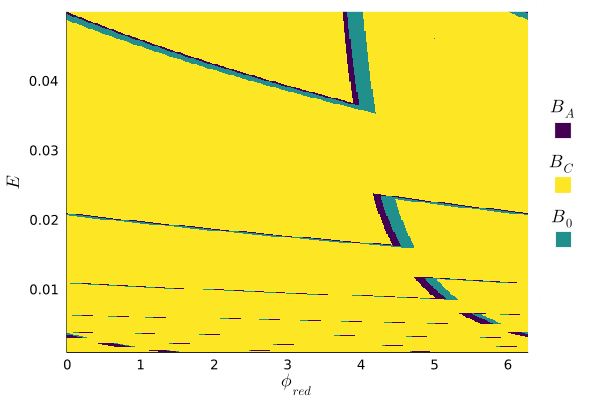}
		\caption{The $\phi_{red}$--$E$ plane for fixed
			$I=0.91$ in the set of initial conditions in channel $A$.
			Each pixel in this plane is coloured according to the
			channel in which the corresponding scattering trajectory
			ends in the far future. Yellow means basin $B_C$, 
			cyan means basin $B_A$ and dark blue means basin $B_0$.}
	\end{figure}



	\section{Conclusions}
	
	We have illustrated our view of the sequential decay with the help of a minimalistic
	model system. And now the reader may ask what are the important properties of a
	model system in order to be able to show the phenomenon of sequential decay. Essential is
	the presence of at least three particles in order to have two fragment and three
	fragment arrangements. And accordingly, we have basin boundaries between two fragment
	and three fragment final asymptotic channels. In such a setting sequential decay occurs for
	initial conditions inside of the basin of a three fragment channel but very close
	to its boundary with two fragment channels as explained in more detail at the end
	of section 3.
	
	In the model system the basin boundaries for energies above the threshold of the
	three fragment channel are smooth because of the absence of topological chaos for 
	positive total energy. For more complicated potentials there exist chaotic invariant
	sets in the phase space also for positive energies ( above the three fragment threshold )
	which may lead to fractal basins and fractal basin boundaries. Then under a change
	of the total energy or some other system parameter we may have such fascinating
	events as basin crises and basin boundary metamorphoses. The interested reader can 
	find detailed discussions of these events in the references \cite{GOY1,AVS,GOY2,ALY,PGL} and also in chapter 5 of \cite{tel}.
	Such complicated boundaries cause additional technical difficulties of our picture of 
	sequential decay but do not cause any changes in the basic view. 
	
	
	In view of our original motivation stemming from quantum systems, a comment on tunnelling 
	is necessary. The originally mentioned decay chains as far as alpha decay and to a
	lesser degree other charged particle decays are concerned, are dominated by the Coulomb 
	barrier which allows extremely long decay times. Indeed for typical reactions including 
	cases of neutron emission in very detailed calculations, no direct 
	three-fragment effects will be necessary to reproduce the experiments up to energies where
	the distorted wave Born approximations explain the phenomena. On the other hand for the
	molecular case, semi-classical approximations combined with the potentials obtained from 
	Born-Oppenheimer microscopic calculations work quite well and in this case orbits near basin
	boundaries may well play a role not yet appreciated.
	
	The famous picture of a wooden billiard with many balls used by Niels
	Bohr,  is often used to show that one ball entering 
	can well give long living excited states in classical mechanics with sequential decay in 
	classical mechanics. This combined with Coulomb barrier effects largely explains the dominance of
	sequential decay in  low energy scattering, at least in nuclear physics except for the smallest
	nuclei.
	
	But as we mentioned already, in the three nucleon system this effect is not dominant, 
	and Fadeev equations have to be used for a more complete understanding. Alt-Grassberger-Sandhas
	equations \cite{alt1,alt} have been applied to few nucleon systems, but even with small nucleon 
	numbers the deviations from sequential decays vanish quite fast, as long as energies are 
	small.

	\section{Acknowledgments}
	
	The authors thank CIC AC -- UNAM, DGAPA for financial support under grant number AG-101122 and CONACyT for financial support under FRONTERAS grant number 425854.

\end{document}